\documentclass[12pt]{iopart}
\begin{document}

\jl{6}

\title{A rotating incompressible perfect fluid space-time}

\author{Zolt\'an Perj\'es\dag, Gyula Fodor\dag, L\'aszl\'o \'A. 
  Gergely\dag\ and Mattias Marklund\ddag}

\address{\dag\ KFKI Research Institute for Particle and Nuclear
  Physics, Budapest 114, P.O.Box 49, H-1525 Hungary}

\address{\ddag\ Department of Plasma Physics, Ume{\aa} University, 
  S-901 87 Ume{\aa}, Sweden}

\begin{abstract}
    A rigidly rotating incompressible perfect fluid solution of
Einstein's gravitational equations is discussed. The Petrov type
is D, and the metric admits a four-parameter isometry group. The
Gaussian curvature of the instantaneous constant-pressure surfaces is
positive and they have two ring-shaped cusps.
\end{abstract}




  Rotating perfect fluid solutions of the field equations of general
relativity have been much sought after because of their importance in
cosmology and in modeling relativistic stars. The purpose of this Letter
is to present a perfect fluid space-time with the metric
\begin{eqnarray}            \label{ds2}
  ds^2 &=& \sin^4\chi (dt+2R\cos\theta d\varphi)^2
     - 2 \sin^2\chi Rd\chi (dt+2R\cos\theta d\varphi) \nonumber \\
     &&- R^2\sin^2\chi(d\theta^2+\sin^2\theta d\varphi^2) \ .
\end{eqnarray}
Here $R$ is a constant. In relativistic units, with Einstein's
gravitational constant chosen $k=1$,
the density is $\mu=6/R^2$. The pressure is a
function of the radial variable $\chi$ alone,
\begin{equation}
p=\frac{4}{R^2} \sin^{-2}\chi-\frac{6}{R^2}\ .
\end{equation}
and the four-velocity has the form
\begin{equation}
u=\sin^{-2}\chi\frac{\partial}{\partial t} \ .
\end{equation}
The pressure is positive in the domain $\chi\in(0,0.95532)$.
The acceleration is $\dot u_adx^a= u_{a;b}u^bdx^a= -2\cot\chi d\chi$,
the shear vanishes and the vorticity vector is parallel to the
acceleration. Hence, in the Collins-White classification \cite{CW} of
perfect fluids, the type is {\sl IIIAGii}. In Herlt's formalism \cite{Herlt}, 
the metric belongs to class I. The weak energy condition is
satisfied since $\mu+p>0$.
 
  To the authors' best knowledge, this is the first example of a rigidly
rotating incompressible perfect fluid space-time. The solution was  
found by Ferwagner \cite{Ferwagner} and studied by one of the authors \cite{Marklund} when investigating 
locally rotationally symmetric perfect fluid space-times of class I 
in the Stewart--Ellis classification \cite{Stewart-Ellis}.
 
  This solution of Einstein's gravitational equations can be obtained
by a procedure similar to the one how the vacuum Kerr space-time arises
from the vacuum Schwarzschild metric in the Eddington form $ds^2=
(1-2mr/\Sigma)dt^2+2dtdr-\Sigma(d\theta^2+\sin^2\theta d\varphi^2)$
(with $\Sigma=r^2$) as the seed metric. One replaces
the coordinate differentials $dt\to dt+a\sin^2\theta d\varphi$ and
$dr\to dr+a\sin^2\theta d\varphi$, leaving $\Sigma$ as a test
potential. The vacuum Einstein
equations then yield $\Sigma=r^2+a^2\cos^2\theta$.
 
  The interior Schwarzschild metric of a static perfect fluid
\begin{equation}
ds^2=(A-\cos\chi)^2dt^2-R^2[d\chi^2+\sin^2\chi
     (d\theta^2+\sin^2\theta d\varphi^2)] \
\end{equation}
is brought to the more familiar form \cite{Schw} by the
coordinate transformation $\sin\chi=r/R$.
The constant $R$ determines the density, $\mu = 3/R^2$. The constant
$A$ is related to the radius $r_1$ of the matter ball by
$A=(1-r_1^2/R^2)^{1/2}$. The four-velocity is
$u=(A-\cos\chi)^{-1}\partial/\partial t$.
 
  There is no coordinate transformation bringing the interior
Schwarzschild metric to a form corresponding to the Eddington metric.
Nevertheless, we find
that the simple metric $ds^2= \sin^4\chi dt^2 -2R\sin^2\chi d\chi dt
-R^2\sin^2\chi(d\theta^2+\sin^2\theta d\varphi^2)$
can be used for generating a perfect fluid space-time. The requisite
substitution here is $dt\to dt+2R\cos\theta d\varphi$ and
$d\chi\to d\chi$.
 
 Choosing a null frame along the principal directions of the curvature,
we get the Weyl spinor component \cite{NP}
\begin{equation}
\Psi_2 = \frac{i\cot{\chi} - 1}{R^2} \ .
\end{equation}
All other Weyl spinor components vanish, thus the Petrov type is {\it
D}. There is a curvature singularity at the center $\chi=0$.
Closed time-like curves exist in the $\partial/\partial\varphi$ direction
in a neighborhood ${\cal N}$ of the axis $\theta=0$, bounded by the 
surface $\tan\theta=2\sin\chi$.
 
  The two-surfaces defined by constant values of the coordinates $t$ and
$\chi$ have a negative definite metric outside the domain ${\cal N}$ and
the pressure is constant on these surfaces. The Gaussian
curvature \begin{equation}
K=4\frac{1+a^2}{a^2R^2}\frac{\sin^4\theta+a^2\cos^4\theta}
                        {(\sin^2\theta-a^2\cos^2\theta)^2}\ ,
\end{equation}
with the constant $a=2\sin\chi$, is positive and has two cusps on the
boundary of ${\cal N}$.
 
Solution of the Killing equations reveals that the metric (\ref{ds2})
contains four Killing vectors:
\begin{eqnarray}
K_1&=&\partial/\partial t       \nonumber\\
K_2&=&\partial/\partial \varphi \nonumber\\
K_3&=&2R\frac{\sin\varphi}{\sin\theta}\frac{\partial}{\partial t}
      +\cos\varphi\frac{\partial}{\partial\theta}
      -\sin\varphi\cot\theta\frac{\partial}{\partial \varphi}\\
K_4&=&2R\frac{\cos\varphi}{\sin\theta}\frac{\partial}{\partial t}
      -\sin\varphi\frac{\partial}{\partial\theta}
      -\cos\varphi\cot\theta\frac{\partial}{\partial \varphi}\ .\nonumber \end{eqnarray}
The symmetry group is $O(3)\times U(1)$. Thus the space-time
(\ref{ds2}) is locally rotationally symmetric and
belongs to category {\sl (i)} of the type-D perfect fluid
classification scheme \cite{Senovilla}.

   The singular behavior of the metric (\ref{ds2}) on the symmetry axis can be
removed on a semiaxis by introducing a new time coordinate 
$t'=t\pm 2R\varphi$. The physical significance of the manifold on wich the 
entire axis is regular, however, is debatable as is the case with the NUT 
space-time.
 
   We obtain a solution of the Einstein equations with a cosmological
term $\Lambda$ by the substitution $\tilde p = p+\Lambda$ and
$\tilde\mu=\mu-\Lambda$. In the cosmological solution, the domain of
positive pressure can be extended or shrunk by a suitable choice of
$\Lambda$.
 
{\ack This research has been supported by OTKA
grants T17176, D23744 and T022563. M M was supported by the Royal 
Swedish Academy of Sciences and the Hungarian Academy of Sciences.}


\section*{References}

\begin{thebibliography}{99}
  \bibitem{CW} Collins C B and White A J 1994 {\it J.\ Math.\ Phys.}\ 
    {\bf25} 1460 
  \bibitem{Herlt} Herlt E 1988
    {\it Gen.\ Rel. Grav.}\ {\bf 20} 635
  \bibitem{Ferwagner} Ferwagner R 1992  
    Solutions of Einstein's field equations for a rigidly rotating 
    perfect fluid, 
    in {\it Relativity Today}, ed.\ Z Perj\'{e}s, 
    (Nova Science Publishers, Inc) 133
  \bibitem{Marklund} Marklund M 1997 {\it Class.\ Quantum Grav.}\ 
    {\bf 14} 1267
  \bibitem{Stewart-Ellis} Stewart J M and Ellis G F R 1968 
    {\it J.\ Math.\ Phys.}\ {\bf 9} 1072
  \bibitem{Schw} Schwarzschild K 1916  
    {\it  Preuss.\ Akad.\ Wiss.}\ 424
  \bibitem{NP} Newman E T and Penrose R 1962 {\it J.\ Math.\ Phys.}\ 
    {\bf 3} 566
  \bibitem{Senovilla} Senovilla J M 1987 {\it Class.\ Quantum Grav.}\ 
    {\bf 4} L115 
\end{thebibliography}
\end{document}